\documentclass{article}

\usepackage{arxiv}
\usepackage{eurosym}
\usepackage[utf8]{inputenc} % allow utf-8 input
\usepackage[T1]{fontenc}    % use 8-bit T1 fonts
\usepackage{hyperref}       % hyperlinks
\usepackage{url}            % simple URL typesetting
\usepackage{booktabs}       % professional-quality tables
\usepackage{amsfonts}       % blackboard math symbols
\usepackage{nicefrac}       % compact symbols for 1/2, etc.
\usepackage{microtype}      % microtypography
\usepackage{lipsum}
\usepackage{graphicx}
\usepackage{color}
\usepackage{amssymb}
\usepackage[ruled,vlined]{algorithm2e}
\usepackage{subcaption}
\usepackage{amsmath}
\usepackage[autostyle]{csquotes}

\usepackage{todonotes}

\usepackage{hyperref}
\usepackage{color}
\usepackage{amssymb}
\usepackage{rotating}
\usepackage[T1]{fontenc}
\usepackage{lscape}
\usepackage[inline]{enumitem}

\title{Marketability of building energy efficiency systems based on behavioral change: A case study of a novel micro-moments based solution}

\author{
  Yassine Himeur\thanks{This paper has been accepted in Fifth International Congress on Information and Communication Technology (ICICT), London, UK, 2020.} , Abdullah Alsalemi, Faycal Bensaali\\
  Department of Electrical Engineering\\
  Qatar University\\
  Doha, Qatar \\
  \texttt{yassine.himeur@qu.edu.qa;a.alsalemi@qu.edu.qa;f.bensaali@qu.edu.qa} \\
   \And
 Abbes Amira \\
  Institute of Artificial Intelligence\\
  De Montfort University\\
  Leicester, United Kingdom \\
  \texttt{abbes.amira@dmu.ac.uk} \\
		\And
Iraklis Varlamis, George Bravos, Christos Sardianos, George Dimitrakopoulos \\
  Department of Informatics and Telematics\\
  Harokopio University of Athens\\
  Athens, Greece\\
  \texttt{varlamis@hua.gr;sardianos@hua.gr;gebravos@gmail.com;gdimitra@hua.gr} \\
}

\begin{document}
\maketitle

\begin{abstract}
In spite of the substantial advance in developing energy-efficient buildings, power demand in the building sector is still remarkably growing due to teleworking and e-learning triggered by the COVID-19 movement restrictions. This is highlighted by the inefficiency of energy saving measures that have recently been set owing to the the marketability failure and unsuccessful design integration of existing ICT based solutions. Specifically, the factors that affect energy efficiency comprise natural and socio-economic restrictions, technological advances, and last but not least the human behavior. Energy efficiency based on behavioral change has attracted an increasing interest in the recent years, unfortunately, solutions in this area suffer from the lack of marketability because of the absence of both prospective studies and consumer awareness. This work, focuses on a key cost-effective technology for monitoring power consumption and for contributing to the behavioral change through smart, personalized, and on the moment, action recommendations. In order to examine the marketability of the proposed solution, we begin with a market and research analysis of the domain of energy efficiency systems in the building sector that target behavioral change. Moving forward, various factors that affect the commercialization of the final product are considered before proceeding with recommended actions to ensure its successful marketability via conducting a Go/No-Go analysis. Finally, based on the comprehensive commercialization study, a GO decision is concluded for the subject technology.
\end{abstract}

% keywords can be removed

\section{Introduction} \label{sec1}
Recent predictions assume that urban population is expected to  double by 2050, leading to an increase of the whole urban power consumption from approximately 240 EJ to more than 730 EJ \cite{carreon2018urban}. Especially in urban areas, only the building sector consumes more than 74\% of the overall energy used in such environments, and accounts for roughly 50--75\% of the CO$_{2}$ emission. This makes buildings the main power consumer and the main reason for gas emissions \cite{HIMEUR2020115872,PAN2020114965}. Further, the consumption could be increased more due some unexpected circumstances. e.g. the COVID-19 pandemic and its impact on energy consumption in households due to the movement restriction, which has promoted teleworking and e-learning, and hence has increased the energy usage in residential buildings. Consequently, developing green buildings including measures to curtail energy usage has become a current challenge, in which governments, decision-makers and utility companies invest large amounts of money every year to develop innovative solutions helping in promoting energy efficiency \cite{HIMEUR2020114877}. 

The market of energy efficiency systems is driven by users' and governments' requirement for high savings' levels that always keep increasing. 
Systems that target residential buildings either incorporate high-efficiency technologies in the design and development of new buildings (e.g. natural heating or cooling solutions) or the retrofitting of existing buildings (e.g. replacement of energy consuming appliances with green ones, use of led lights, etc.) \cite{KARKANIAS20102776, POPESCU2012454}. Human-centered solutions focus on the change of consumers behavior through information, notifications and recommendations. 
Solutions for commercial buildings (e.g. malls, hotels, stadiums, etc.), invest on the direct impact of small changes that scale up to huge benefits (e.g. light bulb replacements, heat pumps, etc.) for the building owners.
Finally, in the case of industrial installations, which accounts for almost a third of energy use, most opportunities exist in the reduction of energy losses through the reuse of power and the heat it generates (combined heat and power - CHP) \cite{york2013frontiers}.
% \todo[inline]{This \cite{york2013frontiers} is a very interesting article to read. Although from 2013, it contains a lot of useful information about the energy saving market.}

%\subsection{Behavioral change as a promising solution to reduce wasted energy}
Recently, energy providers, policy makers and end-users have progressively become aware of the importance of behavioral change towards energy saving and reduction of carbon emission, in residential \cite{diao2017modeling, hu2017survey} and office \cite{galvin2016selling, d2017synthesizing} buildings. In this context, an increasing number of research works \cite{iwasaki2019using,kim2017establishment} and projects \cite{MOBISTYLE2019,casals2020assessing} as well as commercial products \cite{AlertMe2020,Ecoisme2020} have emerged. They all capitalize on the potential of change towards a more energy sustainable behavior, and they address the detection and matching of consumer attitudes with specific actions that must be taken \cite{alsalemi2019endorsing,Sardianos2020IJIS-ERS}.

%\subsection{The role of AI and ICT for promoting energy efficiency}
Many works, take advantage of the widespread use of artificial intelligence (AI), information and communication technologies (ICT) and internet of things (IoT) in order to provide and promote smart solutions that either gradually shape user behavior according to suggested policies \cite{fraternali2017encompass,kar2019revicee,koroleva2019designing} or, even better, deliver the right action at the right moment to the user \cite{sardianos2020rehab}. All these actions have a positive impact on the end-user's awareness about energy efficiency and provide tangible benefits in terms of money savings.

The systems that control the energy intelligent buildings, monitor all the possible aspects of user activity that consume energy, evaluate the impact of all actions and behaviors on the energy saving potential and prioritize the recommended actions accordingly \cite{nguyen2013energy}. In order to maximize recommendations' acceptance, the intelligent systems take into account the elasticity of user needs and estimate the probability of a positive user response to a recommended saving action \cite{ashouri2018development, sardianos2020data}.

%\subsection{Contribution of this paper}
However, before developing and commercializing any new energy efficiency solution, its marketability potential (market assessment) should be evaluated. Explicitly, a thorough analysis of the energy saving market's forces, drivers, barriers, risks, opportunities and potential competitors must be conducted . Therefore, prior to launch a new solution, performing a market assessment is of utmost importance to estimate the potential customer base for it. In this line, a properly implemented marketability study enables us for deciding how to deploy the new solution using our limited resources and further how to pursue the best market opportunities providing the greatest return on investments. By contrast, failing in conducting an accurate marketing assessment may lead to wasting our resources, missing market opportunities, low return of investments and further considerable financial losses.

\begin{figure*}[!t]
\centering
\includegraphics[width=16.5cm, height=6cm]{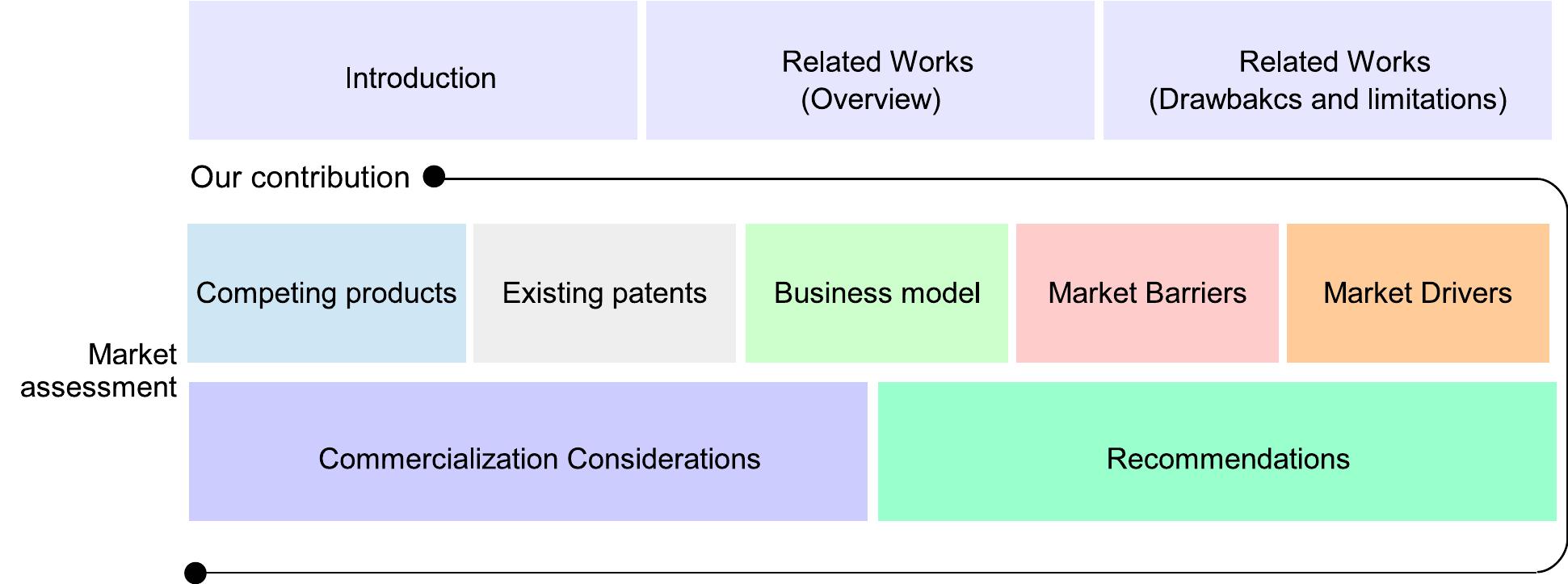}
\caption{Structure and contribution of the proposed framework.}
\label{Cont} 
\end{figure*}

In this paper, we propose to the best of the authors' knowledge the first study that discusses the marketability of energy efficiency systems based on behavioral change with a focus on the micro-moment based solution. In this context, we conduct a comprehensive \textbf{Go/No-Go} analysis based on discussing 
 four areas related to the available products, patents, research projects, and commercialization strategy considerations. In this line, as depicted in Fig. \ref{Cont}, a market assessment study is performed to evaluate the commercial potential of a new energy saving product based on behavioral change and micro-moments.
We also provide overall recommendations based on the factors outlined in the individual sections of this framework. Our recommendations are based on a cursory examination of i) possible competing products, ii) existing/published patents, iii) business model, iv) market barriers, v) market drivers, and vi) commercialization considerations. Moving forward, a set of recommendations are then generated based on factors outlined in the critical analysis conducted through the market assessment.

%\subsection{Organization of the paper}
Section \ref{sec2} that follows, performs a literature review on the related research projects and the frameworks used for the assessment of product marketability. Section \ref{sec3} performs a market overview that begins with energy saving frameworks and projects and continues with the most popular products and the most recent patents on energy efficiency. It discusses their main features and their limitations. Section \ref{sec4} goes deeper in the domain of energy efficiency recommendations. It introduces (EM)$^3$ \footnote{(EM)$^3$: Consumer Engagement Towards Energy Saving Behavior by means of Exploiting Micro Moments and Mobile Recommendation Systems (\url{http://em3.qu.edu.qa/}) }, an energy consumption and activity monitoring system for buildings that brings the marketing-originated concept of \enquote{micro-moments} to the energy efficiency recommendations domain, and analyses its disruptive power. Based on the analysis of the previous sections, Section \ref{sec5} provides some recommendations on the marketing strategy that \enquote{micro-moment} based energy efficiency recommendations must follow and the pitfalls that should be avoided.
Finally, Section \ref{sec6} concludes this work with the next steps on the marketability of the EM3 product.

\section{Literature review} \label{sec2}
The recent advances in energy-efficiency allow modern buildings to save energy by more than 90\% compared to ordinary buildings or even produce energy with energy producing add-ons (e.g. photovoltaic or wind generators) \cite{aydin2019relationship}. However, the access to this kind of buildings is still very limited, especially in developing countries because of the high cost of designing green buildings with structural improvements. Therefore, in addition to the structural changes in the buildings, and the technological update of appliances, the change in user behavior is supported by a large variety of ICT solutions that combines consumption monitoring and recommendations \cite{law2018understanding, jia2017occupancy}. 

In order for an ICT solution to succeed in the market of energy efficiency, it is important to be early assessed on its marketability. The early detection of strengths and weaknesses, and the needs and risks of the market, will allow to  properly introduce and position the product in the market.

\subsection{Frameworks for assessing the marketability of products in the energy sector}
When designing a sustainable marketable product it is important to guarantee its economic, social and environmental aspects \cite{cappa2016deliver}. For technology products, e.g. intelligent systems that combine software with hardware, it is also important to be technologically viable and at least match the  technical features of their main competitors. Monitoring and advisory systems that promote energy efficiency must be offered at a lower price compared to their competitors and must demonstrate higher benefits from energy saving.

According to the net present value (NPV) criterion method \cite{remer1995compendium}, the investment on an innovative project must take into account: i) the free cash flows, ii) the overall costs (including materials' costs, labour and operating costs, and taxation), and iii) the time required for production and deployment. These values have to be projected to the present, future, annual and capitalized worth values in order to test the feasibility and sustainability of the marketed solution.

Due to the importance of energy efficiency systems for the stakeholders of the building sector, including companies, governments and the end-users, a few number of frameworks have been proposed in the literature to assess their marketability. These frameworks begin with the identification of market barriers and drivers. For example, in \cite{Alsop8239226}, the authors conduct a market assessment that aims at informing governments and utility companies where the best energy investments could be achieved for meeting the united nations objectives and improving the living standards of the population in rural regions. Explicitly, different parameters are investigated to identify  the associated strengths and weaknesses of a specific region and which products are more appropriate. While, in \cite{AYDIN2019593}, Aydin et al. focus on studying the relation between building energy saving solutions, their aesthetic characteristics and their marketability. Specifically, the causes of marketability failure are identified before proposing a set of orientations to enhance  the aesthetic features of existing energy efficiency solutions and therefore improving their marketability potentials.

In \cite{BREY20182893}, a marketability study has been conducted for analyzing the cost related to the primary change of tradition gasoline and diesel fuels by hydrogen fuel for road transport in Spain from two points of view: first with reference to the investor active in this sector, and second with regard to the consumer who will use this fuel. 
Moving forward, the authors in \cite{nanduri2012economic} present a bi-level, array game-theoretical model for assessing economic impacts and making operational decisions in carbon-constrained restructured energy-based markets, in which a reinforcement learning scheme has been used that could consider different learning and adaptive factors of the market's participants.

Finally, in the early work of Brambley, et al. \cite{brambley2005advanced} authors performed an assessment of the market for building controls (plant control and maintenance, energy recording and saving) and systems (heating, ventilation and air-conditioning, lighting, security, fire/life safety and access control). The main value proposition they identified comprises: i) enhancement of the indoor environment and build of economic activity, ii) decrease of building maintenance and operations expenses.

\subsection{Limitation and drawbacks}

% \todo[inline]{What do we need here? The limitations and drawbacks of marketability frameworks? Or the limitations and drawbacks of the products they evaluated?}
% \todo[inline]{Re: we can talk on both of them briefly}

The early work of Heinemeier \cite{heinemeier1998marketability} introduced eight categories for assessing the marketability of energy related systems: the system intent, the system value, the action to be taken, the required system reliability, the user notification, the user role, the system cost and the size of the market. These categories influence the methodology for the marketability assessment of any energy related solution. 
The analysis of Aydin, et al on the marketability of energy efficient buildings \cite{aydin2019relationship} identifies the lack of their widespread adoption. It associates it with market failure and applicability problems of the integrated design approach and considers it as a major impediment for the marketability of energy efficiency solutions in buildings. The current lack of an integrated building design that comprises intelligent energy saving solutions, can be an opportunity for future marketability approaches.

Another major drawback of existing approaches is that they leave the human factor out of their scope. Although the aforementioned frameworks have been proposed to assess the marketability of energy saving systems, no work has been dedicated to target the behavioral change based energy efficiency issue. Even when humans are considered part of the energy saving ecosystem, the existing energy efficiency frameworks: i) provide smart visualisations of energy consumption in an attempt to gamify the energy efficiency process \cite{fraternali2017encompass,fraternali2019encompass,koroleva2019designing} but do not help users to perform better with on-time and context aware recommendations, ii) target only the  energy consuming HVAC appliances and do not provide a complete solution for households, commercial or office buildings \cite{wei2018energy}, iii) they provide generic recommendations and tips in order to increase awareness, but do not consider the actual needs and habits of people \cite{paredes2020intellihome} and do not provide eco-friendly alternatives.

The proposed solution covers all the aforementioned limitations in one solution, by combining sensors and smart-plugs, by taking into account users' habitual actions and focusing on the most promising to change and the most beneficial ones (in terms of energy saving), capitalises on the interaction with the users, who take the final decision for an energy saving action or can decide to automate this process. The emphasis in on recommendations' acceptance, so they are persuasive, addressed in the right context and aware of user needs and habits. Visualisation and comparative analytics add to the efficiency of the action recommendations and help the transformation of user habitual behavior.

\section{Market assessment} \label{sec3}
% \subsection{Related projects/frameworks}
The work of Nguyen et al. \cite{nguyen2013energy} offers an interesting survey of energy intelligent buildings based on user activity. The more recent work of Alsalemi et al. \cite{alsalemi2019IEEEsys} focused on the habitual behavior change and surveyed the more recent works.
In Section \ref{sec2} we perform a comprehensive investigation of the state-of-the-art by analyzing a large number of works from different sources, including journals and conference proceedings, white papers, patents, industrial reports and other online sources (e.g. forums, blogs and wikis). In this section, we focus mostly on patents and actual products, since they usually demonstrate higher technology readiness levels (TRL) and are already in (or a step before) the market.

\subsection{Description of existing products} \label{sec3.1}
A search for competitive solutions has been performed using Google Search and a couple of Technology and Market related databases (e.g the EBSCO research databases\footnote{EBESCO: Elton B. Stephens Company, \url{https://www.ebsco.com/corporations/industries/technology-telecommunications}}, InnovationDB products database\footnote{\url{https://www.innovationdb.com/}}, etc.) using terms such as: power,
energy, monitor, monitoring, manage, management, \enquote{energy monitoring}, sensor, software,
network, electricity, micro, moment, \enquote{micro-moment}, and synonyms of these terms, in various
combinations both with and without Boolean connectors.
This search resulted in a long list of products from which we compiled a list of the most relevant and frequently retrieved products. Table \ref{NILMprojects} lists the top products, their manufacturer and their relevance to our project. It also list their application domain and the price, where available.

A large group of commercial solutions, targets public and commercial buildings. Watchwire is EnergyWatch's\footnote{\url{https://energywatch-inc.com/}} cloud based data management, auditing, and reporting platform, that calculates emissions and CO$_{2}$ emissions for commercial buildings. It is used to manage the energy efficiency business process, to simplify energy reporting, reduce energy expenses and increase energy income.
EnergyCAP\footnote{\url{https://www.energycap.com/}} is a family of energy management and energy accounting software products, used for tracking, managing, processing, reporting, benchmarking, and analyzing utility bills and providing energy and sustainability information. It is mostly suited for large organizations with comprehensive energy information management needs.
Energy Manager \footnote{\url{https://www.dudesolutions.com/solutions-energy}} from Dude Solutions analyses utility bills, helps users understanding energy consumption, and identifies cost saving opportunities. The solution is mainly tailored for public and commercial buildings with central utilities.
Finally, GreenerU\footnote{\url{https://www.greeneru.com/}} provides sustainability solutions for Academic buildings targeting centralised utilities (e.g. HVAC optimization, Building automation system, Lighting, etc.).

The second group of products is mainly addressed to households and smaller private or public buildings.
MACH Energy provides a comprehensive suite of energy management software tools\footnote{\url{https://www.machenergy.com/software}}, which are mostly targeted for public and private buildings. MACH's Initiatives automatically identifies energy-saving opportunities, using load profiles of the building and similar buildings in the area and presents facts (savings in dollars, energy consumption, etc.) that help users to prioritize their decisions. SENSE\footnote{\url{https://sense.com/}} is developed around an energy monitor, installed in the household's electric panel. The module processes millions of current readings per second, identifies device signals and presents a whole-home consumption view. It also connects with Google Assistant and Alexa to help users interact with the system. Similarly, the home energy monitor\footnote{\url{https://www.neur.io/energy-monitor/}} by Neurio is installed in the household's load panel and provides granular energy data for devices.

% \todo[inline]{Maybe we can say more about the products of Table \ref{tab:commercial}.}

% \todo[inline]{Re: Yes, we need to add more details and references}

\begin{table} [t!]
\caption{Comparison of commercial solutions for energy efficiency based on behavioral change.}
\label{NILMprojects}
%\resizebox{\textwidth}{!}{
\begin{center}

\begin{tabular}{lllll}
\hline
{\small \textbf{Product name}} & {\small \textbf{Manufacturer}} & {\small \textbf{Relevance}} & {\small %
\textbf{Application}} & {\small \textbf{Price}} \\ 
&  &  &  & {\small \textbf{(USD)}} \\ \hline
{\small WatchWire} & {\small EnergyWatch} & {\small Utility budgeting,
forecasting algorithm, supply market } & {\small Commercial} & - \\ 
&  & {\small projections, delivery tariff rates and public service } & 
{\small buildings} &  \\ 
&  & {\small commission rate cases.} &  &  \\ 
{\small EnergyCAP} & {\small EnergyCAP} & {\small Energy management, data
presentation troubleshooting, } & {\small Work offices,} & - \\ 
&  & {\small and utility bill accounting workflow.} & {\small households} & 
\\ 
{\small Energy Manager} & {\small Dude Solutions} & {\small Dashboard views,
recommendening saving actions and } & {\small Public \& com-} & - \\ 
&  & {\small reporting} & {\small mercial buildings} &  \\ 
{\small GreenerU} & {\small GreenerU} & {\small In-depth analysis and
understanding of campus } & {\small Academic } & - \\ 
&  & {\small energy infrastructure} & {\small buildings} &  \\ 

{\small Energy Management } & {\small MACH Energy} & {\small Actionable
energy data analytics, deep insight into} & {\small Households,} & - \\ 
{\small Software} &  & {\small energy usage and costs, tenant billing
systems.} & {\small public buildings} &  \\ 
{\small SENSE} & {\small SENSE} & {\small Insight into household energy use
and home } & {\small Households} & 300 \\ 
&  & {\small activity through a proprietary iOS, Android, and web apps.} & 
&  \\ 
{\small Home Energy Monitor} & {\small Neurio} & {\small Utility bill
accounting workflow, energy reduction } & {\small Households} & 200 \\ 
&  & {\small tracking, and others} &  &  \\ \hline
\end{tabular}
% \label{tab:commercial}
\end{center}
%}
\end{table}

\subsection{Existing patents} \label{sec3.2}
In order to survey the patent domain of intelligent applications for energy efficiency, we searched the following data sets:
\begin{itemize}
    \item International Patent Documentation (INPADOC) \footnote{\url{https://www.epo.org/searching-for-patents/data/bulk-data-sets/inpadoc.html}}, which contains patents 
    % family documents 
    from 71 world patent signatories and legal status information from 42 patent offices
    \item WIPO Patent Cooperation Treaty database (WIPO-PCT) \footnote{\url{https://www.wipo.int/pct/en/}}, which contains abstracts, full document images, and full text from over a hundred PCT member countries
    \item The European Patent Office (EPO) databases\footnote{\url{https://www.epo.org/}}, which contain Patents and Applications for EU countries.
    \item US Patents and Applications from the US Patent and Trademark Office (USPTO)\footnote{\url{https://www.uspto.gov/}}.
\end{itemize}
 
Using a set of search strings that relate to energy efficiency based on behavioral change, in all databases, we get a list of patents.
The simultaneous search of the patent databases resulted in multiple occurrences of the same patent. This was mainly because a system was registered in more than one databases. We make an assumption that multiple registrations of the same patent in different databases are an indication of
applicants who file, pursue and protect a patent that is more important to its owners than others. Under this assumption, we compiled a list of patents or patent applications which are depicted in Table \ref{NILMpatents}.

These patents and patent applications indicate the research and development (R\&D) direction and drive the state of the art technology in the patent literature. It is worth mentioning that we mainly examine patents from the standpoint of market competition, so it is important for us to identify their specific application environment (e.g. domestic applications, office buildings, etc.)

\begin{table} [t!]
\caption{Comparison of existing energy efficiency based behavioral change patents.}
\begin{center}
\begin{tabular}{llll}
\hline
{\small \textbf{Patent ID}} & {\small \textbf{Relevance}} & {\small \textbf{Environment}} & {\small %
\textbf{Assignee}} \\ \hline
{\small US20100025483A1 \cite{US20100025483A1}} & {\small Collecting sensor-based occupancy and
predicting } & {\small Households,} & {\small Robert Bosch GmbH} \\ 
& {\small consumption } & {\small Office buildings} &  \\ 
{\small US20140099614A1 \cite{US20140099614A1}} & {\small Analyzing user activity data and
identifying deviations} & {\small Households} & {\small Lark Technologies Inc%
} \\ 

%{\small US20120296799A1 \cite{US20120296799A1}} & {\small Monitoring end-users' power consumption
%rewarding} & {\small Households} & {\small -} \\ 

& {\small them for reducing the energy usage} &  &  \\ 
{\small US20140058806A1 \cite{US20140058806A1}} & {\small Energy saving using a network-connected,
multi-} & {\small Households} & {\small Google LLC} \\ 
& {\small sensing learning thermostat} &  &  \\ 
{\small US20170132722A1 \cite{US20170132722A1}} & {\small Promoting energy saving using real-time
social} & {\small Households} & {\small -} \\ 
& {\small energy behavioral change} &  &  \\ 
{\small US20160274556A1 \cite{US20160274556A1}} & {\small Systems and methods for monitoring and
control of } & {\small Households} & {\small Michael D. Murphy} \\ 
& {\small energy consumption systems} &  &  \\ 
{\small US20200034768A1 \cite{US20100286937A1}} & {\small Promoting behavioral change via the
estimation of specific } & {\small Households} & {\small Bizen Green } \\ 
& {\small energy usage and statistical analysis of collected data } &  & 
{\small Energy Corp} \\ 
{\small US9411323B2 \cite{US9411323B2}} & {\small Behavioral energy consumption change and} & 
{\small Households} & {\small LOWFOOT Inc} \\ 
& {\small appliances monitoring using IoT devices} &  &  \\ 
{\small US9927819B2 \cite{US9927819B2}} & {\small Providing end-users with consumption
statistics } & {\small Households} & {\small Honeywell} \\ 
& {\small and controlling of appliances remotely} &  &  \\ 
{\small US20120296799A1 \cite{US20120296799A1}} & {\small System, method and computer program for
energy } & {\small Households} & {\small LOWFOOT Inc} \\ 
& {\small use management and reduction} &  &  \\ 
{\small US9732979B2 \cite{US9732979B2}} & {\small Optimizing energy consumption of HVAC using
behavioral} & {\small Public/domestic } & {\small Google LLC} \\ 
& {\small change and a thermostat} & {\small buildings} &  \\ 
{\small US20110313579A1 \cite{US20110313579A1}} & {\small Energy consumption reducing by comparing
the temporal } & {\small Households} & {\small -} \\ 
& {\small habit pattern with current environmental parameters} &  &  \\ 
\hline
\end{tabular}
\label{NILMpatents}
\end{center}
\end{table}

The conducted patent survey indicates a \textbf{\textit{GO}} in terms of technology. More specifically the patent search shows that home energy monitoring is an area of high interest, with novel solutions, which improve current state of the art, being actively sought. The absence of any single patent or application that clearly combines the same features, functionality, and approach as our proposed technology is an additional indication of its novelty. In combination, these two factors indicate the subject technology may
be of high interest in the market, and may provide clear competitive advantages.

\section{Energy efficiency based on micro-moments} \label{sec4}
Although there are some hardware and software products available for this area,
none appears to be based on the \enquote{micro-moment} approach to monitoring and
analyzing human activity within the home. This may make the subject technology novel
in its approach. Beyond its technical foundation, the subject technology also appears to compare
favorably in terms of features and functionality. It also appears to be expandable to
cover additional parameters such as humidity and temperature.

\subsection{Description of the proposed solution}
\begin{figure*}[!t]
\centering
\includegraphics[width=0.8\textwidth]{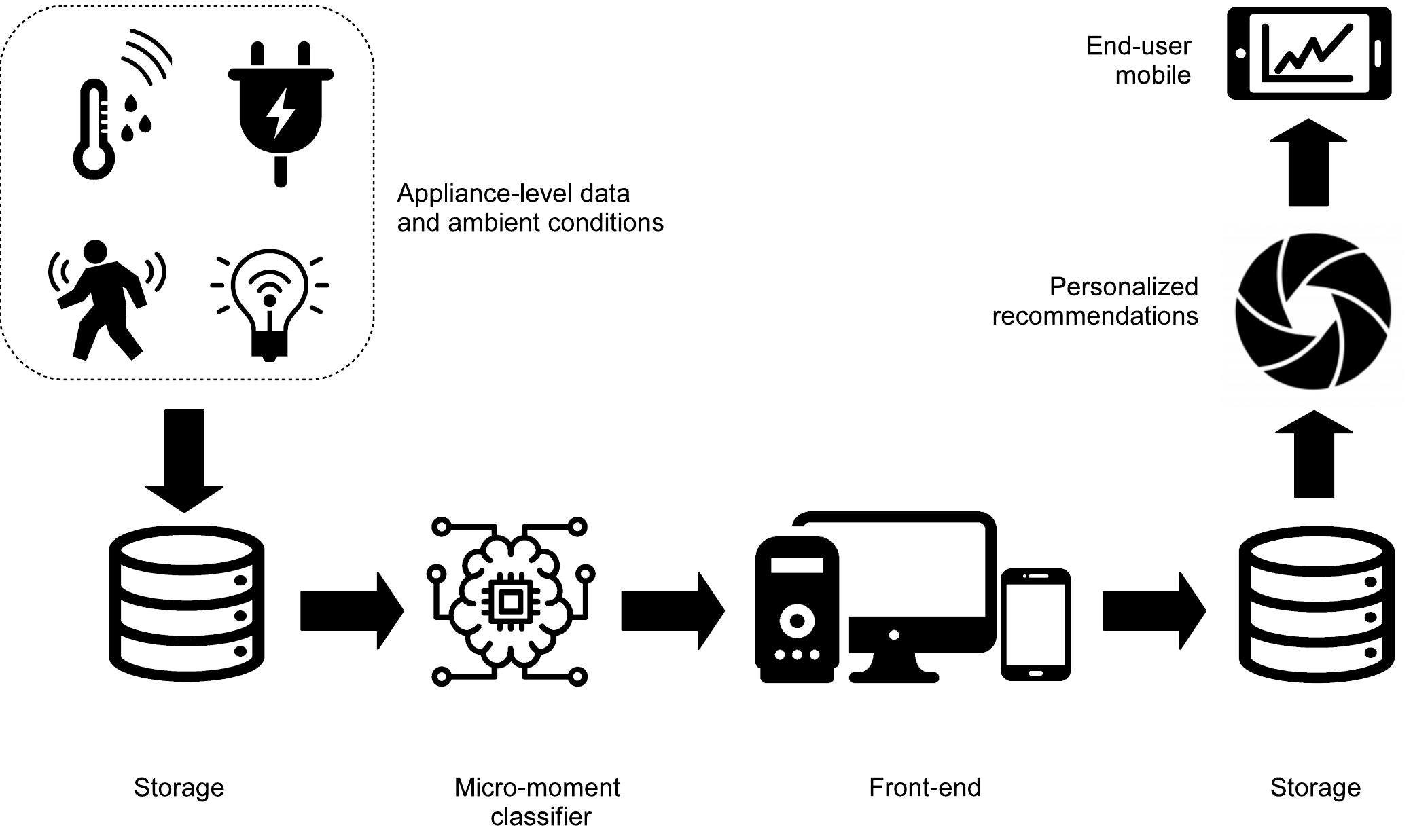}
\caption{Energy saving based on micro-moment behavioral change.}
\label{Micro-moment} 
\end{figure*}

This section presents a detailed description of the (EM)$^3$ based energy saving solution, which is developed for inducing conscious energy consumption behavior. This is possible using a micro-moment analysis that helps in detecting abnormal energy consumption events and a recommender system to provide end-users with tailored and timely advice. \cite{Alsalemi8959214}. Fig. \ref{Micro-moment} illustrates the flowchart of the (EM)$^3$ framework, which encompasses a set of components as follows:

\begin{enumerate}
\item Data collection: gathers electricity consumption footprints, occupancy patterns and ambient conditions. 
\item Micro-moment analysis: identifies abnormal power consumption behaviors using a deep learning model \cite{HimeurCOGN2020,Himeur2020IntelliSys}
\item Recommendation and automation: provides end-users with tailored recommendations for endorsing responsible energy use and possibility for monitoring appliances \cite{Sardianos2020GreenCom}
\item Statistics and visualization: offers end-users their energy consumption statistics in an effective and engaging manner through a mobile application and facilitates the visualization of anomalous consumption data.  
\end{enumerate}

To summarize, the (EM)$^3$ includes tow main parts, smart plug and mobile app. The first one, encompasses sensing devices that are used to capture data, and safely store them in a secure database \cite{AlsalemiRTDPCC2020}. Accordingly, a No-SQL CouchDB server database is utilized for storing data, i.e. end-users' micro-moments and occupancy data, user preferences and habits, power saving recommendations and its acceptance level. 
The (EM)$^3$ smart plug records ambient conditions (i.e. indoor temperature and humidity, room luminosity, and occupancy patterns) and power consumption for a various domestic devices (e.g. light bulb, computer, TV, etc.). It also provides contextual micro-moment information, including operating an appliance, entering/leaving a room, adjusting the settings of an appliance, anomalous consumption of an appliance at a given time, and using an appliance while the end-user is not present \cite{Alsalemi2020ieeeSyst}.

\begin{figure*}[!t]
\centering
\includegraphics[width=0.8\textwidth]{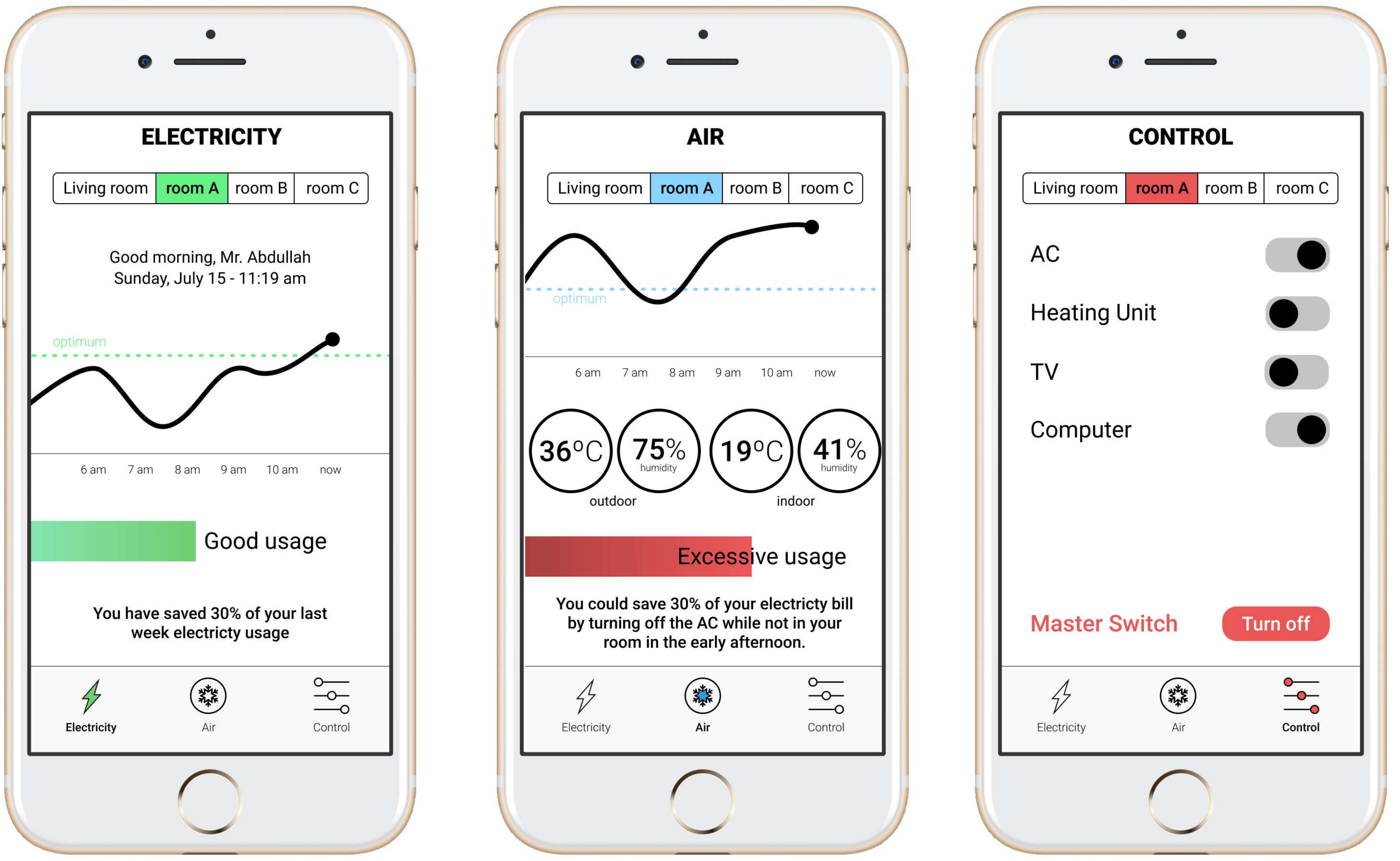}
\caption{Mobile app dedicated to provide the end-user with notifications and recommendations to reduce energy consumption.}
\label{MobileApp} 
\end{figure*}

On the other hand, the (EM)$^3$ mobile app aims to boost energy saving through providing end-users with innovative visualizations of their consumption footprints in real-time. Fig. \ref{MobileApp} portrays the essential screens of the (EM)$^3$ mobile app. Line plots are actually utilized for delivering the information with a small dashboard for presenting more details with reference to the selected screen. Moreover, further energy saving plots are illustrated in addition to ambient conditions (i.e. indoor and outdoor temperature and humidity, indoor luminosity, and room occupancy).

\begin{figure*}[!t]
\centering
\includegraphics[width=16.5cm, height=4.5cm]{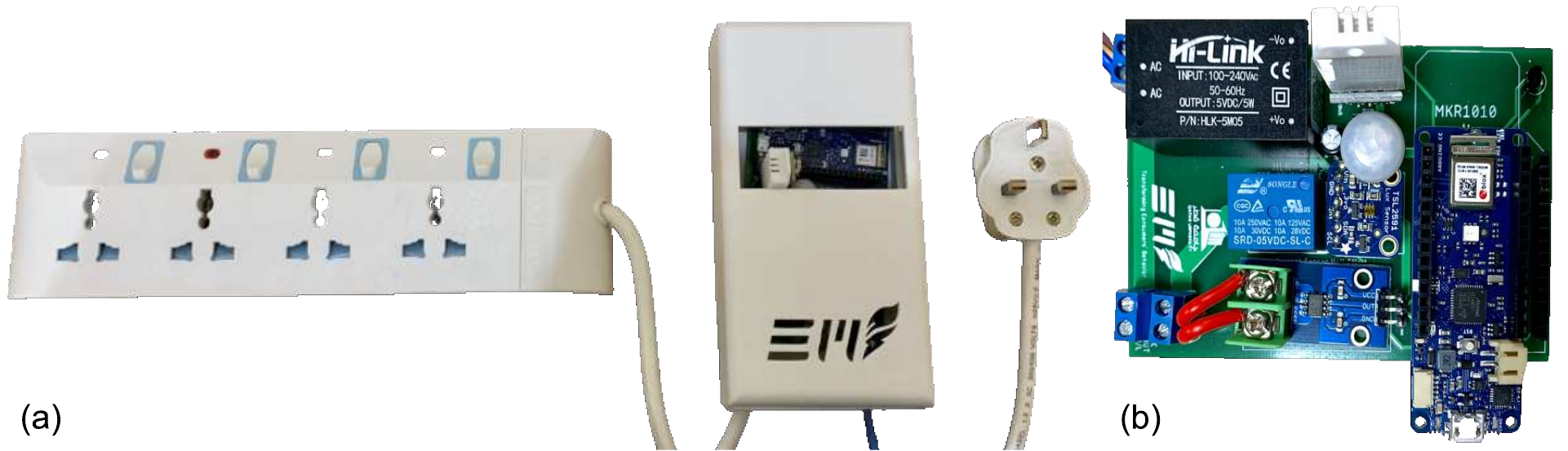}
\caption{Anatomy and PCB of the implemented smart plug.}
\label{smartPlug} 
\end{figure*}

In terms of the hardware implementation, the smart plug encompasses a printed circuit board (PCB), a 3D-printed plastic casing, a plug, and a socket extension as it is illustrated in Fig. \ref{smartPlug} (a). The core of the smart plug is the PCB. The board, shown in Fig. \ref{smartPlug} (b), features a self-powered mechanism through the line of the appliance, eliminating the need of a separate power source to operate it, an occupancy sensor, a luminosity sensor, a temperature and humidity sensor. Also, a relay is included to enable remote appliance operation, in addition, invasive energy monitoring is employed due to the direct connection with the appliance. It is worth noting that the system consumes an average of 45-60mA, which is taken into consideration in analysis and generating recommendations.

The PCB is conceived for accommodating two categories of micro-controllers, which could offer the best compromise between the cost and computing performance. Therefore, it could support both the Arduino MKR-1010 and the ESP32. Specifically, this enables additional testing in terms of performance and wireless capabilities. To provide the reader with more details, we have computed and compared the performance, communication latency, and costs in Table \ref{MicroEM3}.
In terms of the cost of the (EM)$^3$ smart plug, it could be fabricated with a cost ranging between 20 USD - 40 USD according to the used micro-controller board. This is considered as a low cost in comparison to its capability to improving the adoption of residential power consumption monitoring technology.

\begin{table} [t!]
\caption{Performance of the (EM)$^3$ Smart plug per micro-controller}
\label{MicroEM3}
\begin{center}

\begin{tabular}{llll}
\hline
\textbf{User board name} & \textbf{Processing speed (s)} & \textbf{%
Communication latency (s)} & \textbf{Cost (US \$)} \\ \hline
ESP-WROOM-32 & 0.16 & 3.19 & 10 \\ 
Arduino MKR 1010 & 1.05 & 2.25 & 33.90 \\ \hline
\end{tabular}

\end{center}
\end{table}

\subsubsection{Scalability}
In addition to the low-cost property of the (EM)$^3$ solution, the scalability in larger cases is one of the main features of its architecture. In effect, the scalable architecture provides both embedded and external plug and play connectivity. Because every device is managed independently or in combination with other appliances based on the end-user's preferences and habits, the (EM)$^3$ solution could seamlessly scale to larger case scenarios with further appliances and monitoring options needed. Moreover, the requirements needed to implement our solution are not resource demanding and do not depend on the number of observed appliances, and hence it is effortless to expand this application in larger spaces.

\subsubsection{Privacy preservation}

When speaking of marketability of energy efficiency frameworks necessitate a deeper look into the incumbent issues surrounding data management strategies, particularly issues concerned with end-user privacy. 

The growth of cloud computing has allowed unique access to data through possibly secure Internet servers. Many cloud providers have real-time (or nearly real-time) results. They can also include a bundle of additional helpful functionality for end-users to improve the collection and analysis of data. As a result, in the grand scale, cloud computing is on the rise to become the default data storage choice, including energy efficient systems. 

On the other hand, privacy and cyber security problems emerge, particularly in cloud-based solutions. Questions as to whether the data is fully secure on a third-party server can become serious. Also, even though the data is stored on a cloud server operated by the data owner, concerns regarding the reliability of the device can come to light. For e.g., how well is the device secured against cyber security attacks? This survey is of critical significance when working with comprehensive energy use profiles that can be used in an adversely harmful way. 

This is one of the main reasons behind focusing on local data storage disconnected from the Internet and available only to end-users in the "intranet" of the building (e.g. house, education, business). Security is virtually assured in terms of data protection and external threats (except in cases where physical interference is occurring or where the intranet is hacked). All in all, a local storage solution may be deemed plausible if no external internet link is needed. 

However, a device disconnected from the Internet will easily become redundant due to the continuous change in end-user behaviors, behavior patterns and subsequent analyzes of those changes. If the quality of the data does not alter, the retraining of machine learning algorithms with new external data is appropriate. Inevitably, a hybrid model would emerge, providing a balance between local storage and cloud processing. Critical data may be stored on-site on a private cloud server. Open datasets, device algorithms, and user interfaces can be maintained on a cloud platform on the internet.

\subsection{Discussion and Key findings}
These are strong factors that favor the continued development and commercialization of the
subject technology.
We do note a few caveats to consider. For example, although the \enquote{micro-moment} approach
may be novel, the technology still needs to compete in terms of features, functionality, and
price with existing and in-development products. Consumers may not readily understand the
advantages (or even the basic concept) of the micro-moment approach, or how it can be
applied to advantage for home energy monitoring. Instead, they will likely make purchasing
choices based on how effective the technology is and (ultimately) whether or not it effectively
reduces their energy costs. Thus the subject technology will need to be carefully marketed, with
support data presented in easily understandable form, before it is likely to be adopted widely in
the market. Another possible caveat is the lack of IP protection, without it, others may be
able to apply the same approach to develop competing products.

However, these issues can be addressed, giving that the subject technology is developed and performs to expectations. In this regard, the key findings can be summarized as follows:
\begin{itemize}
\item \textbf{Technology's maturity level:} according to setup information of the energy saving based on behavioral change and micro-moments, the subject technology appears to be in the early stages of development, with key components yet to be defined and developed to the working prototype stage.

\item \textbf{Intellectual property:} according to a through investigation of the literature, no patent application has been filed for the technology. Nor have its technical details been publicly
disclosed. Therefor the subject technology may be covered under Trade Secret.

\item \textbf{Possible competing products:} our research suggests there are numerous products, both
hardware and software, designed for monitoring and analyzing energy consumption within
homes and other buildings. These include physical sensors as well as software that collects
and analyzes the data provided by these sensors. However, we found no commercially
available systems that utilize the micro-moment concept to analyze human behavior within
the home/building, and make suggestions based on that data. Nor did we find any system
that clearly provides a more robust feature set. Therefore, the functionality and approach of
the subject technology appear novel and may have commercial application.

\item \textbf{Possible competing patents:} our quick review of recent patenting indicates there is a high
level of activity involved in developing technologies related to home energy monitoring and
management. However, we did not find any single patent or application that clearly
combines the same features, functionality, and approach as the subject technology,
particularly the use of micro-moments. This may indicate the subject technology is novel,
may be of high interest in the market, and may provide clear competitive advantages.

\item \textbf{Possible competing R\&D:} recent publicly disclosed R\&D appears to indicate a moderate to
light level of interest in home energy monitoring, with perhaps 10 to 20 or so recent
projects which could represent likely competition in the near future. For instance, one
energy monitoring system based on detecting and classifying human activities within the
home claims an 18\% reduction in energy consumption. In addition, ongoing public
initiatives such as BENEFFICE \cite{Garbi2019}, ChArGED \cite{Papaioannou2017a,Papaioannou2017b}, OrbEEt \cite{ORBEET2016}, and EnerGAware \cite{casals2020assessing} may represent alternative
solutions for some potential consumers. However, we found no single project based on the
use of micro-moments, nor did we find any project that offers a feature set that compares
favorably with the subject technology.

\item \textbf{Examples of Potential Targets:} A list of potential targets has been described previously in this paper, including SENSE, EnergyWatch, EnergyCAP, MACH Energy, Dude Solutions, among others.

\end{itemize}

\subsection{Business model} 

A proper business model for the proposed micro-moments-based energy efficiency solution elaborates on how such a solution can create and deliver value to the potential stakeholders (prosumers, energy network operators, aggregators etc.), and how they can get remunerated for their engagement. Following the main sections of the business model canvas (BMC) methodology in the following we describe the various aspects of our Business Model. 
Three business models have been identified to be a good fit for the solution, these being software-as-a-service (SaaS), platform-as-a-service (PaaS) and royalty and licensing. The business model outlined below will consider a combination of all three.

The \textbf{Value Proposition} of our solution comprises several benefits for the users. 
The envisioned solutions and products will ensure technological innovation, prosumer-consumer empowerment, energy savings without compromising comfort levels, a balanced mix of all the required actors of the market value-chain, up-to-date service (software) delivery, as well as cost and time effective solution acquisition over the forecast period (i.e. 2020--2030).

% \textbf{Key Partners}
The network of \textbf{Key Partners} comprises the following stakeholders:
% This section defines the network of partners which help to make this business model work. This network comprises the following:
\begin{enumerate*}[label=\roman*)]
    \item the $EM^3$ consortium members,
    \item commercial ICT infrastructure providers,
		\item (EM)$^{3}$ partners,
    \item energy retailers,
    \item energy network operators,
    \item local authorities, and
    \item housing associations.
\end{enumerate*}

The \textbf{Key Activities}, which will ensure that 
% This section outlines how 
the envisioned solution will provide and deliver its value proposition. The Key Activities defined here all help to ensure that the proposed business model can work effectively and efficiently. Main activities include the following:
\begin{enumerate*}[label=\roman*)]
    \item market R\&D, 
    \item evaluation of customer needs, 
    \item assignment of resources, and
    \item marketing.
\end{enumerate*}

The identification of \textbf{Market \& Customer Segments} is important in order to get a better picture of the type of group we are aiming the proposed solutions at. In order to define customer segments for the envisioned solutions, the groups of individuals need to be defined based on their needs, behaviors and other traits they share. The identified actors and stakeholders benefiting from the solutions comprise among others:
\begin{enumerate*}[label=\roman*)]
\item private/public building owners,
\item residential consumers,
\item regional and national companies in the utility and IT, sector in regards to energy solutions for buildings,
\item energy utility companies, aggregators, energy network operators, local service/technology providers, and
\item environmental associations and non governmental organizations (NGOs).
\end{enumerate*}

The envisioned products aim to address a diversified market. The customers we aim to provide these solutions to all have different requirements and needs related to energy efficiency and data visualization. The defined customer segments have few overlaps. However, we see value in investing in all of these diverse segments. The means by which we aim to addresses the specified segments are:
\begin{enumerate*}[label=\roman*)]
\item evidence-based results on the costs and benefits of ICT-enabled energy efficiency techniques,
\item clear and real time guidance,
\item exploiting micro-moments to create recommendation systems,
\item data transparency,
\item usable interface design,
\item ICT Resources escalation,
\item behavioral engineering,
\item adaptive incentivization, and
\item support for exploiting the solution.
\end{enumerate*}

The effective distribution of the solution also depends on the identification of the Distribution Channels, the Customer Relationships and the Revenue streams.

The \textbf{Distribution Channels} can be used to present and promote the solution to the potential customers. It is important to create as many channels and activities as possible in order to effectively spread the message. The main promotional channels comprise: 
\begin{enumerate*}[label=\roman*)]
\item pilot demos, 
\item public consultation \& standardization,
\item publications, info days, dissemination activities,
\item tech Partners \& their resources,
\item website of the project and the application,
\item social media marketing and influence groups, and
\item email campaign tools.
\end{enumerate*}

Various promotional activities can be considered for promoting the developed solution:
% . Some of the main intended are listed below:
\begin{enumerate*}[label=\roman*)]
\item create a pitch deck to promote the solution to potential sponsors and/or investors;
\item set up Google Alerts: Detect where people are talking about the problem the app solves. Also track where people are talking about the app directly by setting up a branded keyword alert;
\item app localization according to market surveys of different focus groups;
\item A/B testing: run pilots of different app versions to different focus groups;
\item search engine optimization;
\item in-campus promotion: directly target audiences with physical means, i.e. billboards, posters in academic and corporate premises;
\item promo video to be used in traditional media and digital (viral) channels;
\item appeal to app review sites to feature the solution;
\item respond to all reviews: provide personalized communication to existing and potential users;
\item apply for awards;
\item write press releases; and
\item write newsletters.
\end{enumerate*}

The \textbf{Customer Relations} define the types of relationships that will be established with the specific customer segments,
% In particular, it is clarified what type of relationship to be established with each customer segment. These relationships are established 
through an array of different channels. In the case of the solutions provided, the partners aim to engage with the users through channels, such as:
\begin{enumerate*}[label=\roman*)]
\item building precompetitive applications - proof of impact and return on investment (ROI), 
\item context-aware triggering,
\item cultural appropriation and localization, and
\item energy solution for policy development.
\end{enumerate*}

The strategic ways in which the project and its partners seek to be paid for the developed solution and their services are depicted in the \textbf{Revenue Streams \& Pricing Models} section of the BMC. Questions that must be answered include:  \begin{enumerate*}[label=\roman*)]
\item what value the customers really are willing to pay, 
\item for what do they currently pay, 
\item how they currently are paying, 
\item how they would prefer to pay, and 
\item how much each Revenue Stream will contribute to overall revenues. 
\end{enumerate*}
The main identified revenue streams comprise:
\begin{enumerate*}[label=\roman*)]
\item smart Energy Dashboard subscription,
\item mobile App revenues (Freemium),
\item customization and professional services,
\item consulting, and
\item white label platform offering (PaaS).
\end{enumerate*}

The \textbf{Key Resources} section defines the necessary resources and skills which help to make this business model and value propositions work. It also emphasizes which distribution channels and customer relationships need to be analyzed in order to realize the proposed revenue streams. In order to interact within the digital world, the envisioned solutions will need to rely on a sound commercial ICT infrastructure. Furthermore, in order to effectively spread the message, it is critical to engage in academic research, labs \& pilots. Of course, when wanting to increase energy efficiency, for any kind of scenario, sound energy consumption data \& models are necessary, these will rely on a real-life customer base and created with smart data analytics tools. In order for a well-suited business model to be established the channels through which customers will be reached need to be discussed and explored, mainly these will encompass business-to-business (B2B), business-to-consumer (B2C) and R\&D pathways.

Finally, the \textbf{Cost Structure} section involves an estimate of most important costs, inherent in the business model. Additionally, it evaluates which Key Resources and which Key Activities are most expensive. Throughout the duration of the project the main costs are identified as:
\begin{enumerate*}[label=\roman*)]
\item market Research,
\item dissemination activities,
\item business model development,
\item deployment,
\item human capital development,
\item platform R\&D infrastructure,
\item community awareness services, and
\item marketing expenses.
\end{enumerate*}
Once developed, the solution will mainly include the costs related to handling customer requests and retaining the audience, as well as costs corresponding to human resources required for the further development and maintenance of the assets.

% \textbf{Revenue Streams \& Pricing Models}
% This section aims to outline the strategic ways in which the project and its partners seek to be paid for the developed solution and their services. Especially for what value the customers really are willing to pay, for what do they currently pay, how they currently are paying, how they would prefer to pay and how much each Revenue Stream will contribute to overall revenues. The main identified revenue streams are outlined below:

% •	Smart Energy Dashboard subscription

% •	Mobile App revenues (Freemium)

% •	Customization and professional services

% •	Consulting

% •	White label platform offering (PaaS)

\begin{figure*}[!t]
\centering
\includegraphics[width=16.5cm, height=10cm]{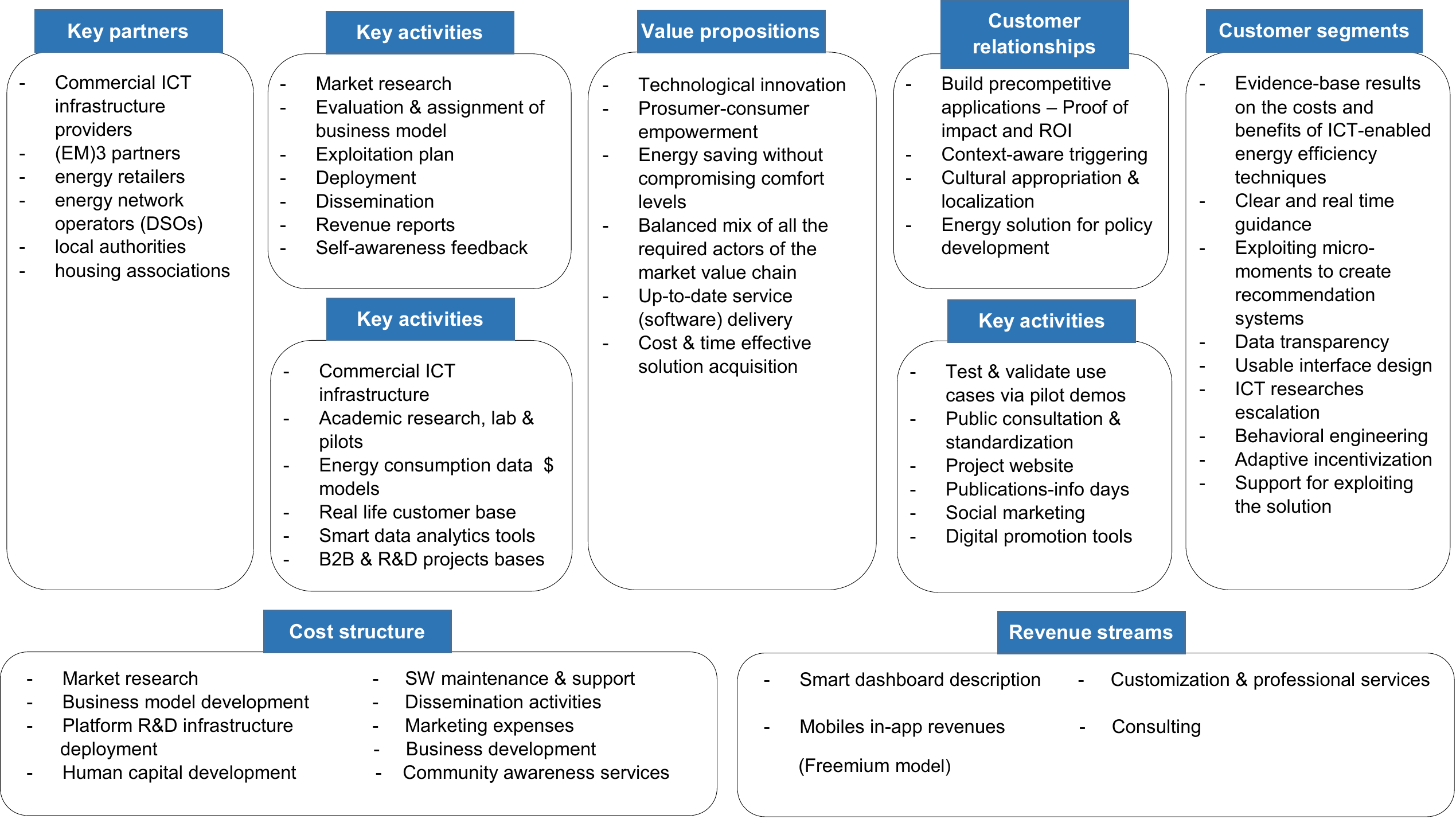}
\caption{BMC describing the key sections for the SaaS approach .}
\label{BM} 
\end{figure*}

The developed assets will be mainly distributed through three methods, namely: i) SaaS, ii) PaaS, and iii) royalty and licensing payments. As mentioned above, for the development of the business model prototype, the BMC was used as guidance. In order to get a more compact view of the identified sections, the partners have combined them into a comprehensive illustration. Fig. \ref{BM} identifies the key sections for the SaaS approach. 

\subsection{Market Barriers}
Various significant barriers, named the market
barriers in state-of-the-art hinder the broad adoption energy efficiency based behavioral change solutions. In our case, we have identify a set of possible market barriers that could be outlined as follows: 
\begin{enumerate}
\item High installation costs accompanied by system complexity are envisaged to raise some difficulties to the growth of the market over the reference time period.

\item A number of major players are already active in the home energy monitoring market.
This includes global companies such as Honeywell International, General Electric,
Comcast Cable (Xfinity), Panasonic, and others. These may comprise a formidable
challenge to new entrants hoping to gain market share.

\item Lack of consumers' awareness concerning home energy management systems (HEMS), behavioral change and the
advantages/profits they offer may impede the marketability of the proposed solution . 
\end{enumerate}

\subsection{Market Drivers}
On the other hand, we have determined an ensemble of possible market drivers, which are defined as the forces that push individuals for purchasing the (EM)$^3$ product and paying for the proposed services. They could be summarized as follows: 
\begin{enumerate}
\item Widespread use of smart-meters, rising investments related to the smart grid technologies, and growing attention to reduce energy cost by efficient deployment of power resources is envisaged to driving the market for the hardware segment over the forecast period, with fastest growth in
the Asia-Pacific region \cite{hossein2020ternet}. 

\item Environmental concerns and increasing energy cost have combined to drive interest in
making more efficient use of energy at home. The necessity for conserving and optimizing energy usage is considered as a key market driver. 

\item It is anticipated that north America will dominate the energy monitoring market during the forecast period. These countries actually concentrate on improving and changing their aging infrastructures, enhancing grid reliability, and allowing smarter electrical networks, which could largely boosting the demand for energy monitoring systems. Furthermore, the U.S. and Canadian electric utilities are intended for investing USD 880 billion and USD 100 billion, respectively, in electrical and energy networks over the upcoming years 2020--2030 \cite{andoni2019blockchain}. These investments would incorporate different sectors, such as smart grid, digitization, energy monitoring, energy management and blockchain among others.

\item Increasing connectivity and widespread endorsement of mobile phones is also expected to positively impacting the energy market growth. 

\item Consciousness raising among energy end-users regarding the durable usage of energy is stimulating demand of energy-efficient devices and HEMS. End-users are understanding that those systems could not only enable to reduce energy expenses, but they are also play a major role in making the existing energy resources
more sustainable. 

\item Demand on HEMS and monitoring devices has assumed greater importance over the last few years due to use of variable pricing approaches provided by service providers. 
Advantageous regulatory policies/initiatives in the U.S. related to energy conservation are intended to incite regional HEMS market growth.

\item Other major factors driving the HEMS market include raising penetration of the internet across both developed and developing economies, increasing role of IoT, big data analysis in energy management, booming market for smart buildings, etc. 

\item Technological development and propagation together with reduced sensor and display costs, improved device-level data processing potential, and roll-out of smart utility meters offer new paths for energy management market growth.
\end{enumerate}

\subsection{Commercialization strategy considerations}
We believe that the development, commercialization, and eventual market introduction should proceed for the subject
technology. One of the first steps could be to seek intellectual property protection in the form of a patent, particularly the part related to its utilization of micro-moments to track and analyze user behavior and activities. However, if protection can be obtained, this may make the subject technology more attractive to potential licensees. It could also make it more difficult to \enquote{reverse engineer} products based on micro-moments. When the subject technology is nearing commercial introduction, a campaign to educate the market should be carefully prepared and undertaken. We feel that users could likely make purchasing
decisions based on features, functionality, and value, rather than novelty of basic
technological approach. Therefore, the performance advantages of the subject technology needs to be identified and emphasized. As noted above, one challenge facing the home energy monitoring market is the lack of awareness and understanding among potential
customers. This challenge needs to be carefully addressed, with users provided a clear
understanding of the subject technology's benefits over currently available options.

Moving forward, the key findings regarding the commercial potential for proposed technology could be summarized as follows:

\begin{itemize}
\item Likely markets and basis for feasibility: home energy monitoring. This is a multi-billion dollar (and growing) global market opportunity, and a number of positive factors appear to be driving future growth.

\item Indicator(s) suggesting how big the market opportunity might be: The global HEMS market was worth US\$ 1.6 Billion in
2018. Looking forward, the market value is projected to reach US\$ 4.4 Billion by 2024, exhibiting a CAGR of around 17\% during 2019-2024.

\item Product opportunities: home energy monitoring and management. Technologies in this niche can likely be adapted fairly seamlessly into the building energy monitoring market, for facilities such as office and municipal buildings, retail stores, and factories.

\end{itemize}

\section{Recommendations} \label{sec5}
There appears to be little question that the home energy monitoring and management market is coming into its own. As discussed in Sections \ref{sec3.1} and \ref{sec3.2}, there are numerous positive factors likely to contribute to continued growth in this arena for the foreseeable future. These include concerns about the environment, the cost of energy, regulatory drivers and initiatives promoting energy efficiency, the proliferation of technical infrastructure on which HEMS can run, and a number of others. Moreover,  two primary challenges in this area; the lack of consumer awareness of these products and their perceived complexity; will likely become lesser factors as home energy systems become more available and adopted.

In this context, before deriving the final recommendations about the commercialization of the proposed (EM)$^3$, we should put more attention to the following factors: 

\begin{itemize}
\item The home power and energy monitoring market, and the building energy market in
general, comprise a multi-billion USD global opportunity. This market appears to be
growing at a robust rate.

\item The drivers for this market appear strong, as consumers are increasingly looking to
make more efficient use of energy and save costs. The market is also being driven by
environmental factors, as consumers are interested in solutions that use less natural
resources without negatively impacting their quality of life.

\item Although there are numerous hardware and software products available for this niche,
none appears to be based on the \enquote{micro-moment} approach to monitoring and
analyzing human activity within the home. This may make the subject technology novel
in its approach.

\end{itemize}

Finally, in order to generate our recommendations, we have evaluated the following aspects; products, patents, research projects and commercialization considerations. For each area, we have provided one of four scores from highest to lowest as below and then an overall score which can be no better than their lowest score of one area. Our summary findings are shown in Table \ref{Go-NoGo}. Based on our commercialization considerations, we conclude GO for the subject technology. As
summarized above, we find a number of strong positive drivers for this market, and challenges
that do not present insurmountable obstacles. We feel this market is favorable for the
introduction of a novel and superior home energy monitoring and management product.

\begin{table}[t!]
\caption{A summary of the signs that we get for commercialization, from the analysis of each factor.}
\label{Go-NoGo}
\begin{center}

\begin{tabular}{ccccc}
\hline
& \textbf{\ \ \ \ \ \ \ \ \ \ \ \ \ Go \ \ \ \ \ \ \ \ \ \ \ \ } & \textbf{\
Conditional Go } & \textbf{Conditional No-Go} & \textbf{\ \ \ \ \ \ \ \ \ \
\ No-Go \ \ \ \ \ \ \ \ \ } \\ \hline
\multicolumn{1}{l}{Product} & Yes & \multicolumn{1}{l}{} & 
\multicolumn{1}{l}{} & \multicolumn{1}{l}{} \\ 
\multicolumn{1}{l}{Patent} & Yes & \multicolumn{1}{l}{} & \multicolumn{1}{l}{
} & \multicolumn{1}{l}{} \\ 
\multicolumn{1}{l}{Research project} & Yes & \multicolumn{1}{l}{} & 
\multicolumn{1}{l}{} & \multicolumn{1}{l}{} \\ 
\multicolumn{1}{l}{Commercialization} & Yes & \multicolumn{1}{l}{} & 
\multicolumn{1}{l}{} & \multicolumn{1}{l}{} \\ 
\multicolumn{1}{l}{Overall} & Yes & \multicolumn{1}{l}{} & 
\multicolumn{1}{l}{} & \multicolumn{1}{l}{} \\ \hline
\end{tabular}

\end{center}
\end{table}

\section{Conclusion} \label{sec6}
Developing energy saving based behavioral change in the building sector along with adopting efficient ICT based systems are the immediate and key solutions to reduce the wasted energy in buildings, especially in unexpected circumstances, e.g. the COVID-19 pandemic situation where the energy demand of buildings is increasing rapidly across the world. Unfortunately, although various research initiative on the application of technology to curtail energy demand have released recently, their implementation in real scenarios is still a paradox due to lack of marketability.

With the aim of bridging the gap between the potential and current accomplished energy saving in buildings, this work performed an overview of the energy efficiency solutions landscape, from research projects to patents and commercial products. Then discussed the key findings of this analysis and formulated the business model of the proposed platform using the Business Model Canvas methodology. By judging all market drivers and impediments and analysing the commercialization strategy considerations, we conclude GO for the subject technology, from all aspects. 

All in all, this framework could be a relevant reference for the energy research community and other researchers from the related topics, especially those developing energy saving solutions and seeking to design their business models in order to improve the marketability of their final products.

\section*{Acknowledgements}\label{acknowledgements}
This paper was made possible by National Priorities Research Program (NPRP) grant No. 10-0130-170288 from the Qatar National Research Fund (a member of Qatar Foundation). The statements made herein are solely the responsibility of the authors.

%\section*{References}

\end{document}